# Density Functional Theory Based Understanding on the Reactivity of $N_2$ Molecule on $Al_n$ (n = 2, 3, 13, 30 and 100) Clusters

**Bhakti S. Kulkarni[†], Sailaja Krishnamurty[\*,‡] and Sourav Pal[\*,†]**

[†]*Physical Chemistry Division, National Chemical Laboratory (CSIR), Pune 411008*

[‡]*Functional Materials Division, Electrochemical Research Institute (CECRI) (CSIR), Karaikudi 630 006*

*Abstract*

Reactivity of Aluminum Clusters has been found to exhibit size sensitive variations. This work is motivated by a recent report[1] predicting higher reactivity of melted Aluminum clusters towards the $N_2$ molecule as compared to the non-melted Al clusters. We attempt to understand the underlying electronic and structural factors influencing the adsorption of $N_2$ molecule (a prerequisite for the reactivity) on ground state geometry (a non-melted structure) of various Al clusters. The results show that the adsorption energy is of the order of 8-10 kcal/mol and does not vary with respect to the cluster size and the electronic properties of the ground state geometry. The structural and electronic properties of high energy conformations of Al clusters (a melted cluster) are also analyzed to explain their higher reactivity towards $N_2$ molecule.

[\*]*Corresponding author, E-mail:* sailaja.raaj@gmail.com; s.pal@ncl.res.in

**I. Introduction**

The appearance of the bulk motif in small sized aluminum clusters has excited many researchers. As a consequence of this property, Al clusters are attracting a lot of attention for their potential applications in optics, medicines,[2] microelectronics[3] and nanocatalysis[4]. In addition to the appearance of bulk motif, ground state $Al_n$ clusters (n < 100) show size specific features in their structures,[5-8] cohesive energies,[9-11] and thermodynamic properties.[12–14] For example, the ground state geometries of many clusters are seen to change from a disordered morphology to an ordered one (or vice-versa) with the addition of a single atom.[15] Adding an extra atom to some clusters also changes the melting transition from a first-order to a second order.[16-22] The above mentioned and many other size dependent characteristics make the study and application of aluminum nano clusters with up to a few hundred atoms both interesting and challenging.

One such application is the synthesis of aluminum nitride. Aluminum nitride, one of the industrially important materials, carries high impact as an electronic material and is usually synthesized through a direct reaction between Al surface or clusters and $N_2$ at a high temperature and pressure[23]. Nevertheless, one needs to modify a processing condition to smoothen such tedious and hard reactions. Consequently, recent experimental studies showing that Al clusters of range 25-100 have melting temperatures that are 450 K, well below the bulk melting temperature, 934 K has excited the several researchers[22]. Not only clusters in this range show a depressed melting temperature, they also show a size sensitive melting behavior. These results can be exploited to design the chemical reactions at desired temperature by choosing an appropriate cluster. In addition, the depressed melting temperature of clusters facilitates easier chemisorption and thus various chemical reactions at very low temperatures (around 450 K). It is noteworthy that $N_2$ molecule is only known to physisorb on the Al surfaces below the melting temperatures[24] and thus use of clusters for such reactions could be more advantageous.

The above aspect is demonstrated very nicely in a recent report by Jarrold *and co-workers*[1] where they discuss the reactivity of $N_2$ on $Al_{100}$ cluster. They have determined the melting temperature of $Al_{100}$ using heat capacity measurements following which the ion beam experiments are used to investigate the reaction between the cluster and molecular $N_2$. They show that above the melting transition, the activation barrier for $N_2$ adsorption decreases nearly by 1 eV. The importance of Al-N reaction has also motivated Romanowski *et. al.*[25] to perform a theoretical study of $N_2$ reaction with liquid Al metal. They have determined the activation barrier for dissociative chemisorption of $N_2$ to be 3.0 eV. They propose that the melting decreases the surface energy, and atoms in liquid are mobile and better able to adjust the $N_2$ molecule. Hence, previous studies on $N_2$ adsorption conclude that the atoms on the surface of the liquid cluster move to minimize their energy lowering the activation barrier.

Apart from the enhanced mobility, very little understanding is available concerning the role of structure and bonding of Al clusters on the adsorption/reactivity of the cluster. The catalytic reactivity is always attributed to specific and precise structural rearrangement of atoms in the material. It is worthwhile to correlate the above two parameters to their reactivity. Thus, the interesting questions are: "Is the chemisorption of $N_2$ molecule a consequence of highly different structure of Al cluster following the phase transition? Do the changes in structure modify the chemical bonding property within the cluster thereby enhancing its reactivity or the higher reactivity is due the dynamical rearrangement of atoms within cluster? Does this reactivity vary as a function of cluster size?" To answer the above questions, we have studied systematically the adsorption behavior of $N_2$ on Al cluster as a function of cluster size. We also address the issue of conformational changes following the phase transition and their impact on $N_2$ adsorption.

To achieve this objective, we choose a series of Al clusters of variable size. It includes low energy conformers of $Al_2$, $Al_3$, $Al_{13}$, $Al_{30}$ and $Al_{100}$ clusters. $Al_2$ and $Al_3$ are the smallest of Al clusters that have been analyzed for understanding the adsorption of $N_2$ molecule. The reason for starting with such small clusters is to have a qualitative understanding on two issues viz., (a) to understand the electronic properties underlying the adsorption of the molecule on the cluster. (b) to analyze the adsorption of $N_2$ molecule as a function of Al cluster size. After analyzing the various reactive sites of these clusters, $N_2$ is adsorbed on them to study its interactive nature. The zero-temperature studies of adsorption is extended to a decahedron conformer as well as a few high energy conformers of $Al_{13}$ (typically seen after the phase transition) to analyze the ease of adsorption on high energy conformations. Since melting enhances distortion, these high energy structures help us to correlate the specific role of structure and bonding towards chemisorption of $N_2$. All considered structures were optimized and the bonding properties within them are analyzed through Electron Localization Function (ELF) and Frontier Molecular Orbital (FMO).

The rest of the work is organized as follows. In Section II, we give a brief description of the computational method and descriptors of reactivity used in this work. Results and discussion are presented in section III for $Al_n$--$N_2$ interaction. Finally, our conclusions are summarized in section IV.

**II. Computational Details**

As mentioned, we have considered aluminum clusters of varying sizes viz.,: $Al_2$, $Al_3$, $Al_{13}$, $Al_{30}$ and $Al_{100}$ for the study. All the possible conformations of these clusters are optimized. Following the optimization, low lying structures are considered for further study of adsorption. All the structures were optimized using Density Functional Theory (DFT) based method. The optimization is carried out using VASP.[26, 27] As in standard DFT programs, the stationary ground state is calculated by solving iterative Kohn-Sham equations.[28] We use Vanderbilt's ultra-soft pseudo-potentials[29] within the Local Density Approximation (LDA) for describing the behavior of core electrons. An energy cut-off of 400 eV is used for the plane wave[30] expansion of Al and N atoms. The structural optimization of all geometries is carried out using the conjugate gradient method,[31] except for the high energy conformers where, we use *quasi-Newton* method[32] in order to retain the local minima. The structure is considered to be optimized when the maximum force on each atom is less than 0.01 eV/Å. Similar optimization procedures are repeated for the interaction of $N_2$ molecule at all Al cluster reactive sites, and for that of bare $N_2$ molecule. All the molecules are enclosed in cubical box of 25 Å dimensions.

Various descriptors are used to analyze the reactivity of the Al clusters. One such is Electron

Localization Function (ELF) modified and applied by Silvi and Savin.[33] According to this description, the molecular space is partitioned into regions or basins of localized electron pairs or attractors. At very low value of ELF all the basins are connected. As this value increases, the basins begin to split and finally, we will get as many basins as the number of atoms. Typically, the existence of an iso-surface in the bonding region between two atoms at a high value of ELF say >0.70, signifies localized bond in that region. The mathematical description of ELF is as follows,

$$\eta(r) = \frac{1}{1+(\frac{D_p}{D_h})^2} \quad (2.1)$$

$$D_h = (\frac{3}{10})(3\pi^2)^{5/3} * \rho^{5/3} \quad (2.2)$$

$$D_p = \frac{\frac{1}{2}\sum_i |\nabla \psi|^2 - \frac{1}{8}|\nabla \rho|^2}{\rho} \quad (2.3)$$

$$\rho = \sum_{i=1}^{N} |\psi(r)|^2 \quad (2.4)$$

where the sum in Eqn. 2.4 is over all the occupied orbitals. $D_h$ is the kinetic energy of the electron gas having the same density or one can say its the value of $D_p$ in a homogeneous electron gas and $\rho$ is the charge density.

We have analyzed the ELF for all clusters and their complexes of $N_2$ at isovalue where atomic basins start merging. Thus, as size and geometry of Al cluster changes, here, the ELF analysis helps us to understand the discrete bonding pattern. In addition, we also analyze the partial charge densities of FMO. The Highest Occupied Molecular Orbital (HOMO) and Lowest Unoccupied Molecular Orbital (LUMO) are the best used global descriptors of reactivity.[34-36a] The $N_2$ adsorption at particular site is followed by HOMO and LUMO analysis, which helps to distinguish the most reactive site. Subsequently, after complex formation, same descriptors elaborate the peculiar Al-N bonding. In addition, to understand the over all charge distribution after complex formation, the charges on various atoms (in optimized bare clusters and respective complexes) are calculated through the GAMESS[36b] considering the single point energy convergence.

Following the studies on the ground state conformations of Al clusters of varying sizes, as mentioned, we have considered few high energy conformations of $Al_{13}$ for $N_2$ adsorption. This is done with an aim of elucidating the role of electronic properties of these high energy melted conformations on the chemisorption. One of the earlier theoretical works has reported that $Al_{13}$ undergoes a solid to liquid transition between 1000-1700 K[37]. Hence, the high energy conformations are obtained from a finite temperature run of $Al_{13}$ at 1200 and 1600K. It may be recalled that an earlier work[37] predicted 1600K to be nearly melting temperature of the cluster. The finite temperature calculations are performed by using Born-Oppenheimer molecular dynamics based on the Kohn-Sham formulation of DFT, employed in VASP. The ionic phase space of the cluster is sampled classically in a canonical ensemble using a method proposed by Nóse[38]. Pursuing the simulation of 10 pico seconds, few conformations were chosen from these high temperature runs. The structure, bonding and reactivity of these clusters are discussed and compared with those of the low lying conformation of the same cluster.

**III. Results and Discussion:**

**(a) Ground State Geometries of Al clusters and their Interaction with $N_2$:**

**(i) Structure, Bonding of $Al_2$ and its Interaction with $N_2$ molecule:**

$Al_2$ like any mono atomic dimer cluster is a covalently bonded one,[39-43] while $Al_3$ is a widely studied system for adsorption of various molecules such as $H_2$[44-46], $O_2$[47] etc. Figure 1 gives the ground state geometries of $Al_2$ and their complexes with $N_2$. The ELF contours and the frontier orbitals for these systems are given in the same Figure. Various inter atomic distances along with the Mulliken charges on each atom [in square bracket] for $Al_2$ and $Al_2$--$N_2$ are given in the same Figure.

We begin with a discussion of $Al_2$ followed by its complex with $N_2$. The inter atomic bond distance in $Al_2$ is 2.59 Å. This is in perfect agreement with the reported literature values.[39-43] The covalent nature of Al$_2$ is well seen from the contours of ELF, where both Al basins merge considerably around a value of 0.76. The FMO analysis predicts that HOMO orbital is bonding orbital (σ-overlap) and LUMO is localized on the two Al atoms (p-orbitals). The favorable mode of N$_2$ interaction with Al$_2$ is a linear structure as seen from Figure 1. Other modes of interaction (perpendicular to the Al-Al bond of Al$_2$ molecule) do not result in local minima. The Al$_2$--N$_2$ complex has an interaction energy of -0.57 eV (-13.10 kcal/mol). Following the adsorption of N$_2$ molecule, the Al-Al and N-N bond distances increase marginally from 2.59 to 2.61 Å and from 1.11 to 1.15 Å, respectively. The Al$_2$--N$_2$ interaction optimizes at an Al-N distance of 1.86 Å. The corresponding ELF contours at an isovalue of 0.76 reveals a polarized electron density on Al(2) and N(1) atoms (see Figure 1). There are no merged basins in the Al-N bonding region even at an isovalue of 0.50.

The HOMO of Al$_2$--N$_2$ complex is composed of 2p orbitals of N atoms while the LUMO is composed of p orbitals of interacting Al atom and N$_2$ atoms. To understand the charge redistribution following the adsorption of N$_2$, we study the difference charge density (Δρ) (Figure 1). The blue region indicates charge gained where as the red region indicates the charge depletion. The presence of red region along the N$_2$ and Al$_2$ bonds and presence of blue region around the each atom shows a small polarization within the whole molecular space. Thus, Al$_2$ and N$_2$, in their interaction show an overall charge transfer from Al$_2$ to N$_2$. In order to quantify the charge transfer we have calculated the Mulliken charges for individual systems and the complex. The charges of Al atoms in the individual Al$_2$ are 6.6*10$^{-4}$ a. u. and -6.6*10$^{-4}$ a. u respectively (and hence nearly zero). The corresponding charges on Al atoms in the complex are 0.123 a. u. (non-interacting Al atom) and 0.513 a. u (interacting Al atom). The N(1) atom (interacting N atom) in N$_2$ shows charge gain of -1.202 a. u. while N(2) is slight positively charged. Thus, both the difference charge density and Mulliken charges reveal an over all charge transfer from Al$_2$ to the N$_2$ molecule.

The moderate interaction energy of N$_2$ with Al dimer promotes us to study the nature of N$_2$ interaction as a function of cluster size and geometry. Does the bonding remain same for larger cluster sizes? Does interaction energy increase with the cluster size? Various such questions will be addressed in the next few sections.

**(ii) Structure, Bonding of $Al_3$ and its Interaction with $N_2$ molecule:**

A three atom cluster has three possible geometric configurations viz., linear conformation, zig-zag conformation and a cyclic conformation.[48, 49] The three conformations in $Al_3$ differ from each other by nearly 1 eV with cyclic $Al_3$ as the most stable configuration. This is consistent with earlier literature reports.[48, 49] Cyclic $Al_3$ is an equilateral triangle as shown in Figure 2 with Al-Al bond distances of 2.47 Å. Thus, we note a decrease in the inter atomic bond distance within the Al atoms as compared to the $Al_2$. In spite of the decreased bond distances, the ELF basins merge only at a value of 0.74 as compared to 0.76 in $Al_2$. The HOMO of $Al_3$ shows a multi centered bonding arising from the overlap of s-p hybridized orbitals. The LUMO is made up of p orbitals of Al atoms. Considering the orbital contributions and their density contours, we verified two modes for $N_2$ adsorption. One perpendicular to the plane of triangle and another along one of the Al-Al bonds as shown in Figure 2. The first $Al_3$--$N_2$ conformation was found to be a meta stable one with $N_2$ getting desorbed. However, the second $Al_3$--$N_2$ configuration resulted in a stable minima with an interaction energy of -0.47 eV (i. e. -10.77 kcal/mol). Interestingly this is nearly 3 kcal/mol lower than that of dimer complex.

$N_2$ adsorption reduces the symmetry of the $Al_3$ cluster from an equilateral triangle to an nearly isosceles triangle. The Al(2)-Al(3) (see Figure 2) increases to 2.64 from 2.47 Å with rest of the two Al-Al bond lengths remaining around their original values. The Al-N interaction converges to the same bond length as that in $Al_2$--$N_2$ complex i. e., 1.86 Å. The $N_2$ bond also increases to 1.15 Å from 1.11 Å like in the case of $Al_2$-$N_2$. The ELF contour of the complex (at an isovalue of 0.80) shows polarization of densities on Al(3) atom (the Al atom interacting with $N_2$) and N(1) atom. Similar to $Al_2$--$N_2$ complex, $Al_3$--$N_2$ complex does not show merged basins in the Al-N bonding region. The frontier orbital analysis shows that HOMO is contributed by the p orbitals of $N_2$, Al(3) and Al(2) atoms. The LUMO is composed of bonding orbital of Al(1) and Al(2) atoms (the two Al atoms not interacting with $N_2$). The charges derived by Mulliken population analysis for $Al_3$ and $Al_3$--$N_2$ are given in Figure 2. All the atoms in the $Al_3$ are nearly neutral in the cluster as well as the complex. However, the $N_2$ molecule is polarized following the complexation. However, the difference charge density plot shown in Figure 2 reveals that there has been some charge redistribution among the Al and N atoms. Each atom is seen to donate some electrons from one of its orbital and gain in another of its atomic orbital there by neutralizing the overall charge transfer. Thus, $Al_3$, shows a marginal decrease in the interaction energy.

**(iii) Structure, Bonding of $Al_{13}$ and its Interaction with $N_2$ molecule:**

Our next cluster to be studied for $N_2$ adsorption is an $Al_{13}$ cluster. $Al_{13}$ is the most well studied among the aluminum clusters. Icosahedra ($I_h$) is indebted global minima in all extensive static and dynamic studies of $Al_{13}$.[50] Hence, we found it interesting to chose this as the next larger cluster for $N_2$ adsorption. Figure 3a shows optimized geometries of $I_h$, and its $N_2$ complex. We now classify atoms within the cluster in order to facilitate the discussion of various reactive sites with the cluster and the $N_2$ adsorption on them. The Various atomic sites in the $I_h$ conformation can be classified into three types viz., "A", "B" and "C" depending upon their chemical environment and their distances from the central atom (see Figure 3a). Site A is the single central atom. Site B and C are the surface atoms containing six atoms each. The atoms "B" lie at a distance of 2.66 Å from the central atom "A" and atoms "C" lie at a distance from 2.60 Å from central atom "A". Figure 3b shows another perspective of the $I_h$ conformation for a better understanding of the reactive sites. Thus, the structure is highly symmetric. Table 1 gives details of these inter atomic distances between various types of sites. Distances between two adjacent "B" atoms is 2.82 Å, while that of two adjacent "C" atoms is 2.90 Å. Owing to their different orientations with respect to the central atom "A", the inter atomic distance between two diagonally placed "B" atoms is 5.32 Å while the distance between "C" atoms is 5.21 Å. The inter-atomic distance between site "B" and "C" is 2.70 Å.

This symmetry is reflected in Mulliken atomic charges (tabulated in Table 1) and ELF isosurfaces (seen in Figure 3a). The site A, is positively charged with 3.43 a. u., sites "B" and "C" attracting a negative charge of -0.3 and -0.28 a. u., respectively. Analysis of ELF basins shows that ELF basins do not merge until a isovalue of 0.75. Around 0.74, basins corresponding to sites "B" and "C" merge with each other. Interestingly, basins of adjacent "B" do not merge with each other. Same is the case for adjacent "C" atoms. These basins merge at a slightly higher value of 0.72. These results indicate an overall covalent nature of the bonds (as in case of $Al_2$ and $Al_3$ clusters), with stronger sharing between the negatively charged "B" atoms with negatively charged "C" atoms. These bonding peculiarities of $I_h$ geometry are also reproduced in the FMO's. Analysis of HOMO reveals a multi centered bonding contributed by "C" atoms in the same plane as seen in Figure 3a. However, the LUMO is contributed by only "B" atoms.

Following the analysis of FMO and ELF, site "B" and "C" appear to be susceptible[51] to the $N_2$ attack. Hence, we have studied the adsorption of $N_2$ on both these sites. The adsorption of $N_2$ on site "B" results in a meta-stable structure with finally $N_2$, desorbing from it. The adsorption of $N_2$ on site "C" results on the other hand in a stable conformation with an interaction energy of -0.34 eV (-7.91 kcal/mol). Interestingly, this is nearly half of that found for $O_2$ molecule on $Al_{13}$ (-0.77 eV) reported by Shiv Khanna and co-workers[47]. They attribute such a low interaction energy of $Al_{13}$ to its magic cluster property (exceptional stability) and hence its nonreactive behavior towards many reagents. The marginally lower interaction energy of $Al_{13}$ towards $N_2$ molecule as compared to $Al_2$ and $Al_3$ is reflected in its Al-N bond distance (which is 1.90 Å as compared to 1.86 Å in $Al_2$ and $Al_3$ complex). The N-N bond distance in the complex is 1.13 Å as compared to 1.15 Å in $Al_2$ and $Al_3$.

Following the $N_2$ adsorption there is a small loss of symmetry. For the sake of better understanding of structural changes following the adsorption we rename the interacting "C" atom as "$C_{int}$". The "A"-"C" bond distance in the $Al_{13}(I_h)$--$N_2$ complex reduces to 2.54 Å from 2.60 Å, "A"-"$C_{int}$" increases to 2.63 Å from 2.60 Å. Similarly, "B"-"C" bonds increase to 2.78 Å from 2.70 Å. The "A"-"B" and "B"-"B" bond distances in the complex remain around their original values. The distance between "C"-"C" is 2.86 Å in the complex as compared to "C"-"C" distance of 2.90 Å in $I_h$. The interatomic distances between the diagonally opposite "B" atoms in complex is 5.50 Å (as compared to 5.32 Å) and that between diagonally opposite "C" atoms is 5.28 Å (as compared to 5.21 Å). Thus, the volume of the cluster increases marginally following the $N_2$ adsorption.

Analysis of FMO's shows that HOMO is contributed by the interacting Al atom, and some lower plane of atoms in the cluster. The charge distribution in $Al_{13}$--$N_2$ complex is given in Table 1. There is an overall charge transfer of 0.21 electrons to $N_2$ from $Al_{13}$ atom. It may be noted that this is less than that seen in $Al_2$ but more than what is seen for $Al_3$. The Al atom bonded to the $N_2$ molecule acquires a positive charge of 0.81 a. u. As a consequence, following the charge transfer (see Figure 3c), the atoms on upper half (atoms above the central atom "A") of $Al_{13}$ are more negatively charged compared to those on lower half (atoms below the central atom "A"). The finer details of the charge distribution (see Figure 3c) show that the positive charge on central atom "A" increases marginally by 0.05 a. u.

The ELF of $Al_{13}$--$N_2$ complex is shown in Figure 3a. It is interesting to note that there is lesser polarization of electron density around $N_2$ as compared to the case of $Al_2$ and $Al_3$. This and the fact that the first merging of the basins occurs only at 0.76 isovalue reflects the lower interaction between the $Al_{13}$ and $N_2$.

**(iv) Structure, Bonding of $Al_{30}$ and its Interaction with $N_2$ molecule:**

To evaluate the adsorption energy of $N_2$ molecule as a function of cluster size, we next consider a 30 atom cluster. There have been quite a few recent reports on the lowest energy conformation of $Al_{30}$. One of the reports has suggested a double tetrahedron as a local minima for $Al_{30}$.[52] Recent reports, on the other hand have reported few other conformations as the lowest energy geometry for $Al_{30}$.[53, 54] However, we have used the double tetrahedron for the case of $N_2$ adsorption. The geometries of $Al_{30}$ and its complex with $N_2$ and various other properties are given in Figure 4. Based on symmetry, the atoms in the cluster can be classified into 7 types of sites as shown in Figure 4. The inter-atomic distances between various sites are reported in Table 2. It is seen from the Table that the edge atoms are far from central atom ("E"). These bonds lay in the range 3.78 Å to 6.47 Å. On the other hand, surface atoms F and G lie at optimal distance of 2.75 Å from central atom while their corresponding distances with adjacent surface and edge atoms is slightly longer. The bond lengths among edge atoms viz., "A"-"B" (2.65 Å), "B"-"C" (2.65 Å) and "C"-"D" (2.59 Å) are the shortest ones.

The short bond distance between the edge atoms is well reflected in the ELF plots (see Figure 4). At an isosurface of 0.86 the basins of all edge atoms merge. However, at the same value, the basins on surface atoms remain as such which merge at a lower value of 0.74. The HOMO of this conformation is mainly contributed by "A" and "B" atoms. The LUMO is contributed by "C" and "D". Considering the FMO contributions, $N_2$ can preferably adsorb on the edge atoms "A", "B" "C", and "D" atoms.

First, we consider site "A" for $N_2$ adsorption (see Figure 4). The Al-N and N-N bonds optimize to 1.86 Å and 1.13 Å respectively as in the case of $Al_{13}$. The interaction energy falls in the same range as observed for the earlier clusters, i. e. 8.01 kcal/mol. The structural details of $Al_{30}$-$N_2$ complex are reported in Table 2. Although, the original double tetrahedron is not distorted much, there are some small perturbations following the $N_2$ adsorption. Notable among them are the inter-atomic distances between site "C" and "D" which increases and consequently, ELF at 0.86 show some disconnected basins. The basin around site A atom, involved in Al-N bond, disappears. The HOMO of the complex is formed of P-P orbital overlap of site C, site D edge and site F, site G surface atoms. However, the LUMO is similar to the LUMO of bare $Al_{30}$. In addition to this, LUMO is also present on $N_2$ molecule.

In the above $Al_{30}$--$N_2$ complex, the $N_2$ is adsorbed vertically on the site A. Adsorption of $N_2$ molecule on sites B and C result in similar interaction energies and electronic properties. We have also attempted a case of multiple bonding where $N_2$ molecule adsorbs on site B with slight inclination towards site C, forming a bridge. However, this orientation results into a conformation where $N_2$ adsorbs only on site B atom with more or less same interaction energy.

**(v) Structure, Bonding of $Al_{100}$ and its Interaction with $N_2$ molecule:**

We must not forget that the original work reported by Jarrold *et. al.* is on $Al_{100}$ cluster. Hence, to check the $N_2$ interaction with such a large cluster, we consider one of the potential minima of $Al_{100}$ cluster. It is difficult to illustrate the in detail structural parameters of 100 atom clusters, hence, we only outline the peculiarities found in bonding those obtained from ELF and FMO reported in Figure 5. The Figure 5 shows that most of the Al atoms are clustered on the surface and very few of them form the core. The covalency of $Al_{100}$ cluster is as good as that of symmetrical $Al_{30}$ cluster. The basins on the surface atoms start merging at an 0.84 isovalue. The HOMO, is centered on many surface atoms, particularly showing the multi-centered bonding resembling that of Ih HOMO. The $N_2$ is adsorbed on one of the free sites. The interaction energy is 7 kcal/mol with an optimized Al-N bond of 1.91 Å. The N-N elongation, here, is to the same magnitude of 1.13 Å. Similar to small sized Al cluster $N_2$ complex, ELF does not show and merged basins along Al-N bond. The ELF contours merge at lower isovalue of 0.80 resulting in further reduction of covalency. The HOMO of $Al_{100}$--$N_2$ is localized on both N atoms and interacting Al atom. On the contrary, $N_2$ also participates in LUMO formation. Thus, the interaction energy of $N_2$ molecule with the ground state conformation does not seem to vary substantially with respect to the cluster size.

**(b) Structure, Bonding and Interaction of High Energy $Al_{13}$ conformations with $N_2$ Molecule:**

**(i) Decehedron ($D_h$):**

As discussed in Section I, $N_2$ adsorption on Al clusters increases dramatically above temperatures which correspond to the phase transition of the later. Just below and above the phase transition, clusters visit several high energy conformations. Such high energy conformations exhibit contrastingly different structural and electronic properties (particularly the surface atoms) as compared to the ground state conformations. One of the objectives of this work is to understand if such a change in surface properties can contribute to better adsorption of $N_2$ (and hence better reactivity) on Al clusters. Hence, we next study some of the high energy conformations of Al clusters and $N_2$ adsorption on them. For this purpose, we have chosen high energy conformations of $Al_{13}$ as a case study. It has been reported earlier that $Al_{13}$ undergoes a phase transition around 1400 K.[37] Decahedron is a dominant high energy conformation of $Al_{13}$ seen from 400 K to 1200 K.[37] Hence, we found it interesting to study the adsorption of $N_2$ on this conformation. In addition, we have also chosen two high energy conformations from a finite temperature run of 1600 K. In the next few sections, we discuss the structure and bonding of these high energy conformations and their implications on the $N_2$ adsorption.

We begin with a discussion on decahedron ($D_h$). Figure 6 shows the structural details of this conformation. $D_h$ having lower symmetry than that of $I_h$, the atoms within it can be classified into 5 types. Figure 6 shows five unique sites (as compared to three seen in $I_h$). The central atom, "A" is bonded to two vertex atoms "B" with an interatomic distance of 2.66 Å. Rest of the ten atoms form two planes between the central atom and two vertex atoms and are classified as "C", "D" and "E" based on the symmetry (See Figure 6). The inter atomic distances between various sites are tabulated in Table 3. Interestingly, the inter atomic distances between sites "E" and "D" (i. e. "D"-"D" and "E"-"E" bond distances) are 2.59 Å, same as that of $Al_2$ dimer. Such a short bond distance is indicative of covalent bonding between these atoms. "C"-"C" bonds are the next shortest ones (2.63 Å) Thus, intra-planar atoms are bonded to each other with strong covalent bonds. The inter-planar bond distances viz., "C"-"D" and "D"-"E" are marginally longer with a reasonably large bond distances of 2.70 Å and 2.81 Å, respectively. Rest of the bond distances lay in the order of 2.66-2.72 Å. Analysis of ELF basins reveals the covalent nature of bonding across the two planes. The ELF basins of same sites merge at a high isovalue of 0.90, whereas those across the two planes merge at 0.78 isovalue.(see Figure 6). Thus, $D_h$ conformation is a more covalently bonded conformation as compared to the ground state $I_h$ conformation of $Al_{13}$. The Mulliken charges on each site are given in Table 3. All the surface atoms are negatively charged resulting in a positively charged central atom. The atoms in the two planes are the most negatively charged ones.

The HOMO and LUMO of $D_h$ are reported in the Figure 6. The HOMO contributed by the bonding orbitals of p-p overlap between the "C"-"C", "D"-"D" and "E"-"E" atoms while LUMO is concentrated only on site B i. e. vertex atom. Considering the contribution of various atoms to HOMO and LUMO, we finalize four sites for $N_2$ interaction, viz., "B", "C", "D" and "E". Interestingly, sites "B" and "C" do not show any reactivity towards $N_2$ adsorption. All orientations of $N_2$ on these two sites resulted in high energy structures with $N_2$ getting finally separated. The presence of highly covalent bonds around site "C" may be a probable reason for desorption on that site, site "B" being least negatively charged may be unfavorable for the $N_2$ adsorption. However, sites "D" and "E" are better for $N_2$ adsorption leading to stable complexes. The site E gives elevated interaction energy compared to site D. In both these optimizations although we started with initial $D_h$ geometry, the final complex obtained was nearly $I_h$--$N_2$ complex. That is during optimization $D_h$ structure transforms into an $I_h$ structure. This has been found to be the case irrespective of the optimization algorithm and initial orientations of $N_2$ molecule on these sites. Thus, $N_2$ adsorbed and stabilized complex is no more $D_h$--$N_2$ complex but it is same as original $I_h$--$N_2$ complex. This change of geometry reflects in ELF and FMO plots shown in Figure 6 which shows that the contours of this complex are analogous to those in original $I_h$--$N_2$ complex. However, the strength of interaction at both sites, D and E, is different. Site E shows almost double interaction energy than that of site D, The interaction energy at site E is 15.86 kcal/mol and the same at site D is 8.28 kcal/mol. These high energies of interaction are obtained from using definition of interaction energy as follows,

$$\text{Interaction Energy} = E_{complex} - (E_1 + E_2) \qquad (1)$$

Where, $E_{complex}$ is energy of complex which is converted to $N_2$ adsorbed $I_h$ geometry. $E_1$ is energy of optimized $D_h$ and $E_2$ is energy of optimized $N_2$. Since, final complex is no longer $D_h$, if we substitute $E_1$ as energy of optimized $I_h$, both interaction energies (at site D and site E) decrease significantly. In the later case site D is almost not bonded to $N_2$ (IE ~ 0.0 kcal) where as interaction of site E and $N_2$ is just same as original $I_h$--$N_2$ complex. As a consequence of this $D_h$ turned $I_h$--$N_2$ complex shows similar behavior of bonding as described above in $I_h$ section. Rest of the geometrical parameters of this final complex are same as those found in $I_h$--$N_2$ complex.

**(ii) High Energy Conformations H1 and H2:**

We next attempt the adsorption of $N_2$ molecule on some of the high energy $Al_{13}$ conformations which are not so well known. As mentioned earlier, these conformations have been extracted from an molecular dynamical simulation of the $Al_{13}$--$I_h$ cluster around 1400 K. Figure 7 gives one of the perspectives of both these two clusters. It is clearly seen that these high energy conformations are distorted leading to as many chemically unequal sites. H1 has nine chemically (Site "A" to Site "I" as shown in Figure 7) unequal sites while H2 has as many sites as the number of atoms within it. The high energy conformation H1 appears to be a hybrid of $I_h$ and $D_h$ conformations. The ELF at an isosurface of 0.90 reveals that both these conformations are characterized by the presence of covalent bonds at few pockets of the clusters. The inter atomic distances between the atoms in the two clusters range from 2.54 to 2.84 Å. The charge distribution as obtained from the Mulliken population analysis for both the conformations is given in Table 4. It is seen from the Table that the atoms on the surface are mostly negatively charged and the central atoms are positively charged. Upon adsorption of $N_2$, H1 conformations results in a stable complex with an adsorption energy of 7.65 kcal/mol while adsorption of $N_2$ on H2 conformations leads to an interaction energy of 13.54 kcal/mol. Interaction energy increases by approximately 5 kcal/mol for a high melted and distorted structure H2, however, the case is not same for H1--$N_2$ complex.

Thus, interestingly it appears that varying the cluster size and just the reorientation of the bonds does not contribute significantly to the better adsorption of $N_2$ on Al clusters. The clear understanding of this elevated strength of interaction in case of H2 conformer can be understood with keen observation of the FMOs. The HOMO (see Figure 7) is dominantly spread over Al (7). In addition this is the least bonded and hence free site. The $N_2$ is adsorbed in a plane of triangle made up of Al(5), Al(7) and Al(12) atoms. This forms a stable complex with Al-N bond optimized to 1.90 Å. Here, the increase in interaction energy however is not ascribed by nature of H2 directly but is approved by additional Al-N bond formed between Al(12) and N(1). The characteristic bond formed so is 2.93 Å. The N-N elongation is 1.13 Å. The Al(7)-Al(12) bond in bare H2 is 2.58 Å. In the respective complex, the Al(12) is close to Al(7). This bond reduces to 2.55 Å as $N_2$ approaches to Al(7) and thus retains a favorable geometry to form another Al-N bond (see Figure 7, H2--$N_2$ HOMO). There is no specific trend i. e. bond elongation or reduction observed in H2--$N_2$ complex formation. In general, the bonds in close vicinity of Al(7) show bond reduction. We also summarize the atomic charges on the interacting Al and N atoms. Let us concentrate on the atomic charges which participate in Al-N bond. A positive charge of 0.38 a .u. is built on the Al(7). The Al(12) atom is also bonded with N(1). Hence, the Al(12) pulls charge of -0.34 a. u. from N(1). The N(1) here, hence, acquires slightly less negative charge compared to other $N_2$ complexes.

Thus after studying various conformers of $Al_{13}$ and their interaction with $N_2$, we can conclude that in case of $Al_{13}$, the $N_2$ interaction is not just dependent on structural arrangement of atom but is also a function of multiple bonding of $N_2$ with Al cluster. In other words, the highly melted or deformed structure where adjacent Al-Al bonds are not so strongly coordinated, $N_2$ can interact to form various Al-N bonds resulting in highly stable $(Al)_n$--$N_2$ complex. This is validated by the analysis of average interatomic Al-Al distances in $Al_{13}$ clusters. The average interatomic distance shows a substantial increase in H2, a highly melted structure (3.91) as compared to 3.59, 3.65 and 3.70 in Ih, Dh and H1 respectively. Here, we must address that $N_2$ adsorption is carried out on a melted structure obtained after phase transition and no temperature effect is consider while its interaction. Thus, to count its chemisorption as mentioned by Jarrold *et. al.* we must bring in the thermal effects with the help of dynamics. Hence, in our next paper, we plan to study the high temperature thermodynamics to obtain highly melted 100 atom cluster, followed by the room temperature reaction of $N_2$ adsorption.

**IV. Conclusion and Scope**

Density Functional Calculations have been carried out to understand the adsorption of the $N_2$ molecule on the ground state geometries of $Al_n$ clusters (n=2, 3, 13, 30, 100). The studies were also extended to few high energy conformations of $Al_{13}$. We summarize the adsorption energies of $N_2$ molecule on these ground state geometries and high energy conformations of $Al_{13}$ clusters in Figure 8. With the exception of $Al_2$, all other ground state conformations show interaction energy between 8-10 kcal/mol. As seen from the figure, the interaction energy is stable with respect to the cluster size. This interaction is mostly accompanied by a charge transfer from Al atoms to the nitrogen atoms in the complex. The bonding within ground state geometries in all cluster sizes is partially covalent and partially metallic. However, the extent of covalancy is seen to vary to some extent between the various sized clusters. Thus, independent of the electronic and structural properties, $N_2$ interacts weakly (an ionic bond) with the ground state geometries of Al clusters. The presence of high energy conformations of $Al_{13}$ on the $N_2$ adsorption is seen to give mixed results with some high energy conformations leading to similar interaction energies as that of a ground state conformation (Decahedron and H1), whereas, high energy conformation of $Al_{13}$, viz., H2 is seen to favor the $N_2$ adsorption better due to the presence of more than one Al-N bond. Such multiple bonds between the ligand molecule and cluster are, thus, a characterstic phenomenon of high temperatures. This is a consequence of increased interatomic distances between the Al-Al atoms (and thus weaker bonds) following the melting. Thus, the above work indicates that the enhancement of reactivity of melted Al clusters is more due to therodynamic factors and electronic factors play a minor role in the increasing adsorption energy of $N_2$ molecule.

**Acknowledgments**

One of the authors (BSK) acknowledges CSIR (Council of Scientific and Industrial Research) for funding of the SRF (Senior Research Fellowship). We acknowledge the Center of Excellence in Scientific Computing at NCL. SP acknowledges the J. C. Bose Fellowship grant of DST towards partial fulfillment of this work.

**Table 1: Structural and Electronic Properties of $Al_{13}$ ($I_h$) and its $N_2$ Complex**

| Distance (Å) | $I_h$ | $I_h$--$N_2$ |
| --- | --- | --- |
| A-$C_{int}$ | - | 2.63 |
| A-B | 2.66 | 2.65 |
| A-C | 2.60 | 2.54 |
| Al-N | - | 1.90 |
| N-N | 1.11 | 1.13 |
| B-C | 2.70 | 2.78 |
| B-B (diagonally placed) | 5.32 | 5.50 |
| B-B | 2.82 | 2.85 |
| C-C (diagonally placed) | 5.21 | 5.28 |
| C-C | 2.90 | 2.86 |
| **Interaction energy** (kcal/mol) | - | 7.91 |
| **Average Charges on various sites as obtained from Mulliken population analysis (a. u.)** | | |
| Sites | $I_h$ | $I_h$--$N_2$ |
| A | 3.43 | 3.49 |
| B | -0.3 | -0.51, -0.52, -0.49, -0.30, -0.30, -0.27 |
| $C_{int}$ | -0.3 | 0.81 |
| C | -0.28 | -0.49, -0.36, -0.41, -0.28, -0.26 |
| N(1) | - | -0.22 |
| N(2) | - | 0.1 |

**Table 2: Structural and Electronic Properties of $Al_{30}$ and its $N_2$ Complex**

| Distance (Å) | $Al_{30}$ | $Al_{30}$--$N_2$ |
|:---:|:---:|:---:|
| A-B | 2.66 | 2.54 |
| A-E | 6.47 | 6.00 |
| B-B | 2.86 | 2.91 |
| B-C | 2.65 | 2.58 |
| B-F | 2.76 | 2.69 |
| B-E | 4.69 | 4.60 |
| C-C | 2.68 | 2.70 |
| C-F | 2.68 | 2.74 |
| C-D | 2.59 | 2.63 |
| C-G | 2.74 | 2.64 |
| F-G | 2.64 | 2.62 |
| C-E | 3.78 | 3.90 |
| F-E | 2.75 | 2.84 |
| D-G | 2.65 | 2.68 |
| G-G | 2.52 | 2.65 |
| G-E | 2.75 | 2.80 |
| D-E | 4.51 | 4.77 |
| Al-N | - | 1.86 |
| N-N | 1.11 | 1.13 |
| **Interaction energy** (kcal/mol) | - | 8.01 |

24Table 3: Structural and Electronic Properties of Al$_{13}$ (D$_h$) and its N$_2$ Complex

| Distance (Å) | D$_h$ | D$_h$--N$_2$ (I$_h$--N$_2$) |
|---|---|---|
| A-B | 2.66 | 2.66 |
| A-C | 2.66 | 2.62 |
| A-D | 2.69 | 2.66 |
| A-E | 2.69 | 2.53 |
| Al-N | - | 1.90 |
| N-N | 1.11 | 1.13 |
| B-C | 2.69 | 2.79 |
| B-D | 2.72 | 2.73 |
| B-E | 2.67 | 2.70 |
| B-B | 5.32 | 5.28 |
| C-D | 2.70 | 2.78 |
| C-E | 4.42 | 4.31 |
| C-C | 2.63 | 2.78 |
| D-E | 2.81 | 2.68 |
| D-D | 2.59 | 2.77 |
| E-E | 2.59 | 2.74 |
| **Interaction energy** (kcal/mol) | - | 15.86 |

Average Charges on various sites as obtained from Mulliken population analysis (a. u.)

| Sites | D$_h$ | D$_h$--N$_2$ (I$_h$--N$_2$) |
|---|---|---|
| A | 2.93 | 3.49 |
| B | -0.10 | -0.31, -0.51 |
| C | -0.30 | -0.26, -0.35, -0.30, -0.30 |
| D | -0.26 | -0.50, -0.34 |
| E | -0.24 | 0.77, -0.40 |
| N(1) | - | -0.21 |
| N(2) | - | 0.10 |

Table 4: Electronic Properties of H1, H2 ($Al_{13}$) and its $N_2$ Complex

| Sites | H1 | H1--$N_2$ |
|---|---|---|
| **Average Charges on various sites as obtained from Mulliken population analysis (a. u.)** | | |
| A | 2.07 | 1.61 |
| B | -0.16 | 1.46 |
| C | -0.01 | -0.18 |
| D | -0.20 | -0.25 |
| E | -0.36 | -0.26 |
| F | -0.28 | 0.00 |
| G | -0.37 | -0.43 |
| H | -0.07 | -0.25 |
| I | 0.01 | 0.00 |
| N(1) | - | -0.22 |
| N(2) | - | 0.09 |
| **Interaction energy** (kcal/mol) | - | 7.65 |

| Sites | H2 | H2--$N_2$ |
|---|---|---|
| **Average Charges on various sites as obtained from Mulliken population analysis (a. u.)** | | |
| 1 | 1.22 | 1.20 |
| 2 | -0.17 | -0.38 |
| 3 | 0.11 | 0.22 |
| 4 | -0.53 | -0.46 |
| 5 | 0.34 | 0.17 |
| 6 | -0.14 | -0.34 |
| 7 | -0.49 | 0.38 |
| 8 | -0.13 | -0.19 |
| 9 | -0.54 | -0.42 |
| 10 | 0.17 | -0.02 |
| 11 | 0.42 | 0.44 |
| 12 | -0.26 | -0.34 |
| N(1) | - | -0.18 |
| N(2) | - | 0.09 |
| **Interaction energy** (kcal/mol) | - | 13.54 |

**Figure 1: Al$_2$ and its N$_2$ interaction from ELF and FMO**

| Property | | ELF | HOMO | LUMO |
|---|---|---|---|---|
| Al$_2$ | 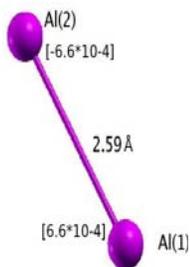 | 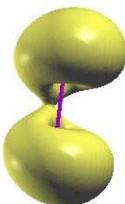 (iso-value=0.78) | 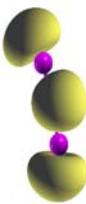 (1/3$^{rd}$ of total charge density) | 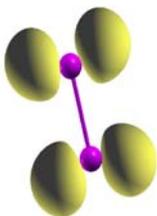 (1/3$^{rd}$ of total charge density) |
| Al$_2$--N$_2$ | 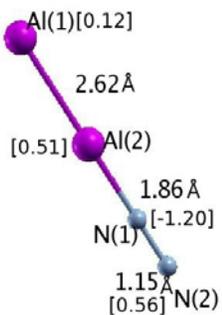 | 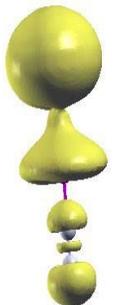 (iso-value=0.76) | 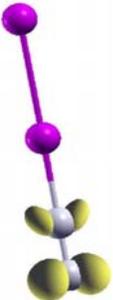 (1/3$^{rd}$ of total charge density) | 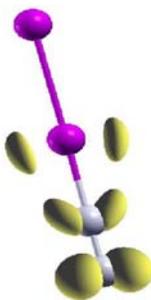 (1/3$^{rd}$ of total charge density) |
| Difference Charge Density | 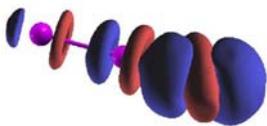 | | | |

**Figure 2: Al$_3$ and its N$_2$ interaction from ELF and FMO**

| Property | | ELF | HOMO | LUMO |
|---|---|---|---|---|
| Al$_3$ | 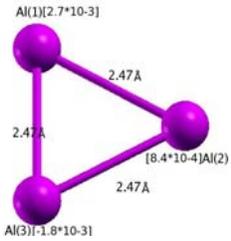 | 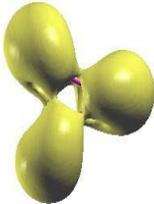 (iso-value=0.74) | 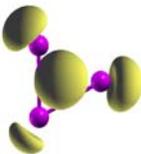 (1/3$^{rd}$ of total charge density) | 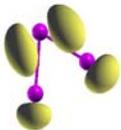 (1/3$^{rd}$ of total charge density) |
| Al$_3$--N$_2$ | 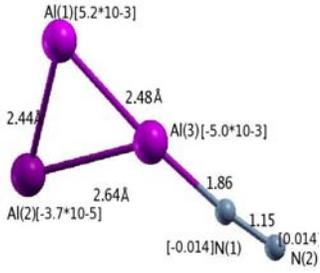 | 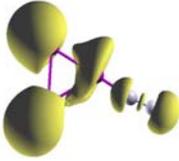 (iso-value=0.80) | 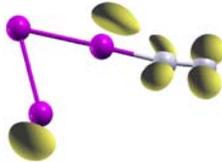 (1/3$^{rd}$ of total charge density) | 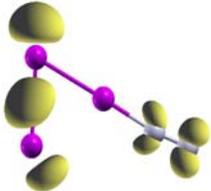 (1/3$^{rd}$ of total charge density) |
| Difference Charge Density | 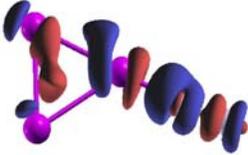 | | | |

**Figure 3a: $Al_{13}$ ($I_h$) and its $N_2$ interaction from ELF and FMO**

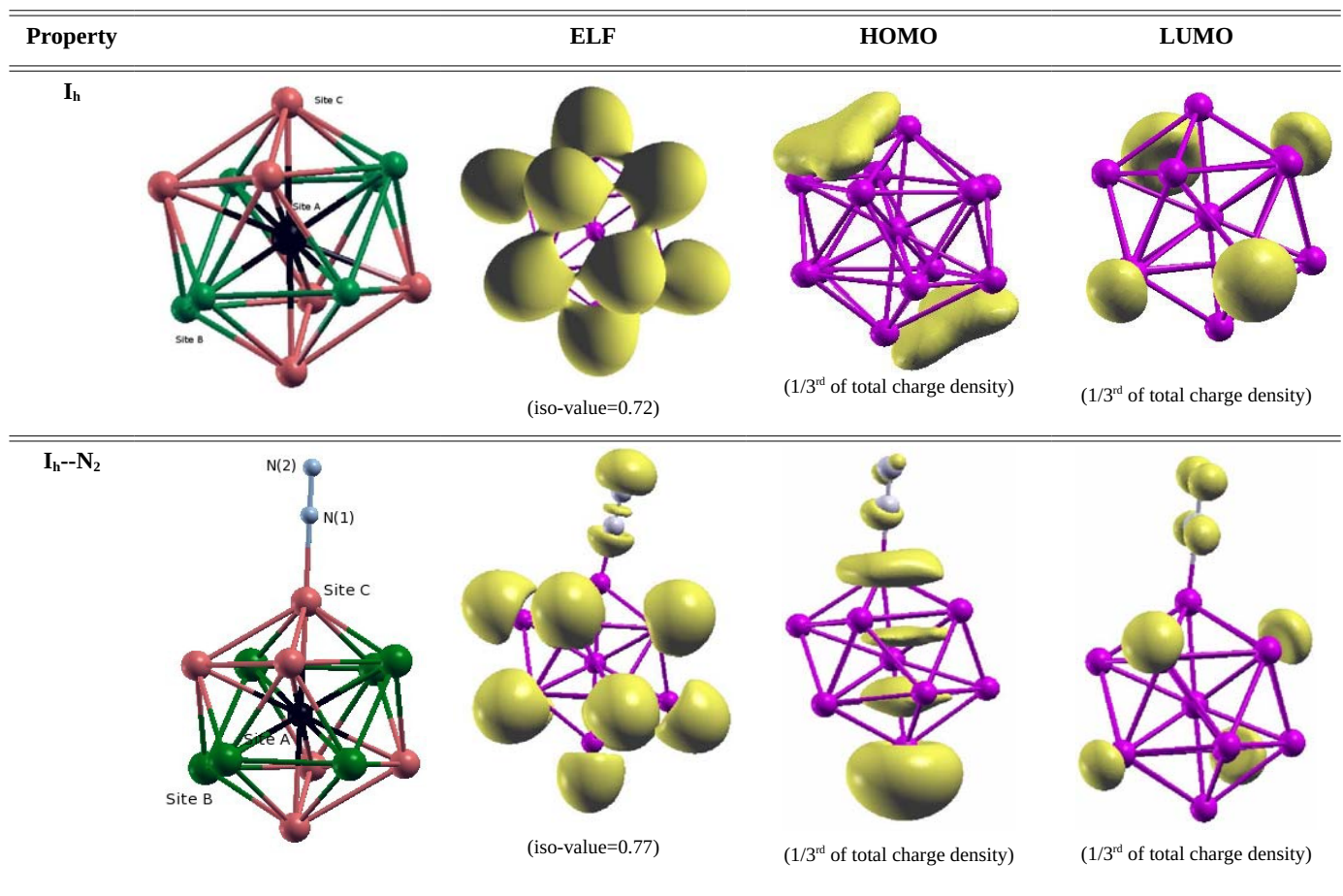

| Property | ELF | HOMO | LUMO |
|---|---|---|---|
| $I_h$ | (iso-value=0.72) | (1/3$^{rd}$ of total charge density) | (1/3$^{rd}$ of total charge density) |
| $I_h$--$N_2$ | (iso-value=0.77) | (1/3$^{rd}$ of total charge density) | (1/3$^{rd}$ of total charge density) |

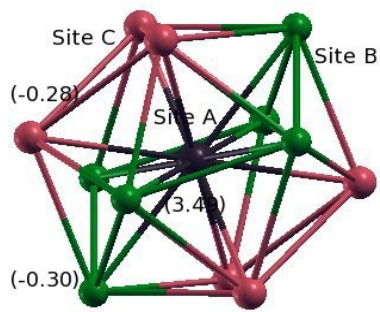

**Figure 3b: $Al_{13}$ ($I_h$) another perspective.**

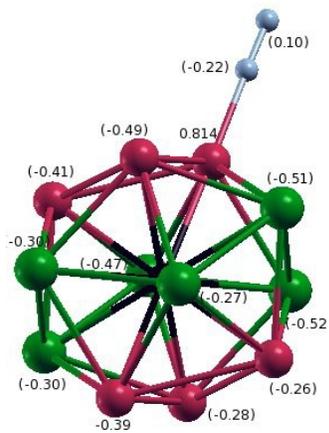

**Figure 3c: $Al_{13}$ ($I_h$--$N_2$) another perspective.**

**Figure 4: Al$_{30}$ and its N$_2$ interaction from ELF and FMO**

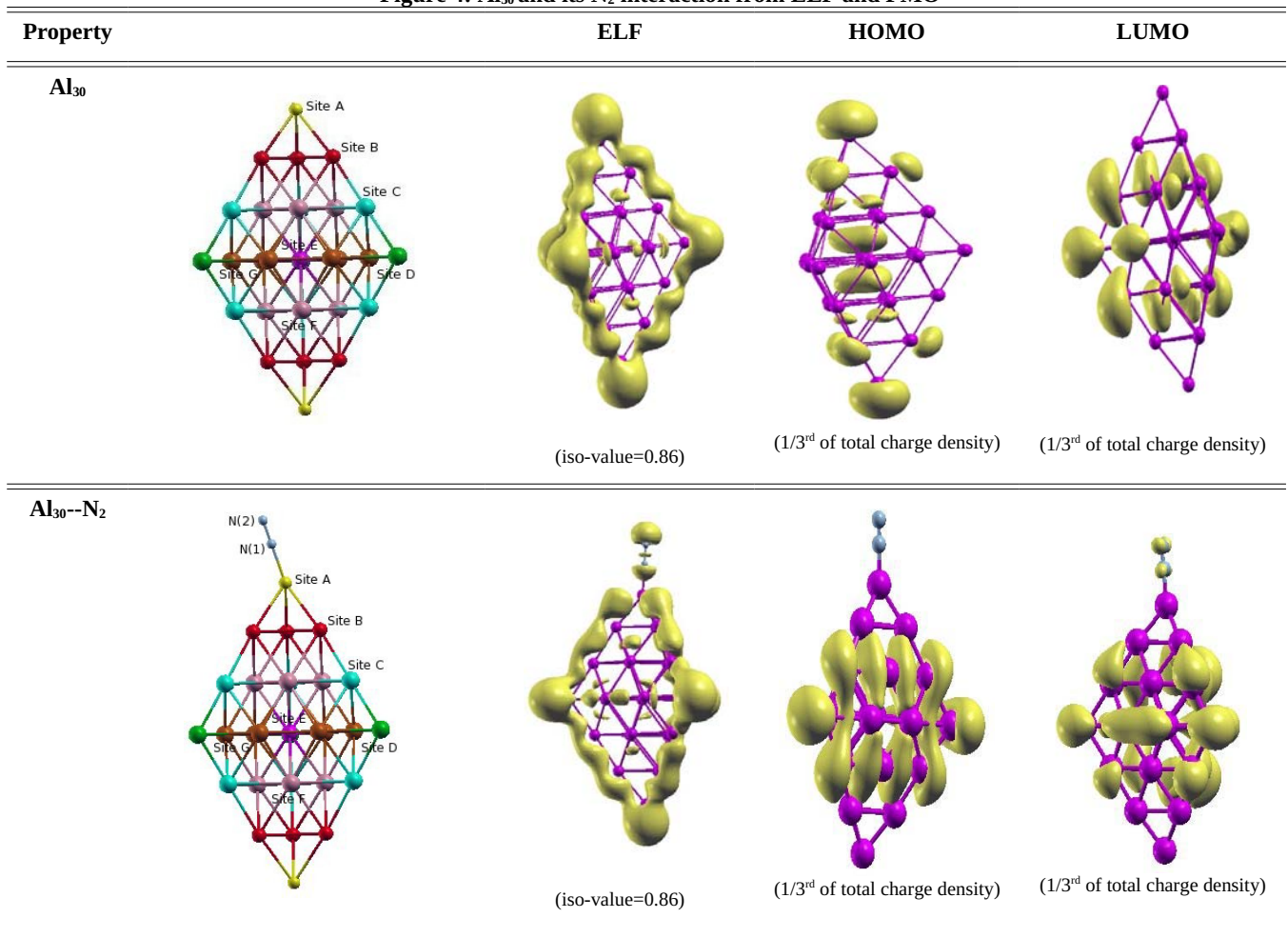

| Property | | ELF | HOMO | LUMO |
|---|---|---|---|---|
| Al$_{30}$ | | (iso-value=0.86) | (1/3$^{rd}$ of total charge density) | (1/3$^{rd}$ of total charge density) |
| Al$_{30}$--N$_2$ | | (iso-value=0.86) | (1/3$^{rd}$ of total charge density) | (1/3$^{rd}$ of total charge density) |

**Figure 5: Al$_{100}$ and its N$_2$ interaction from ELF and FMO**

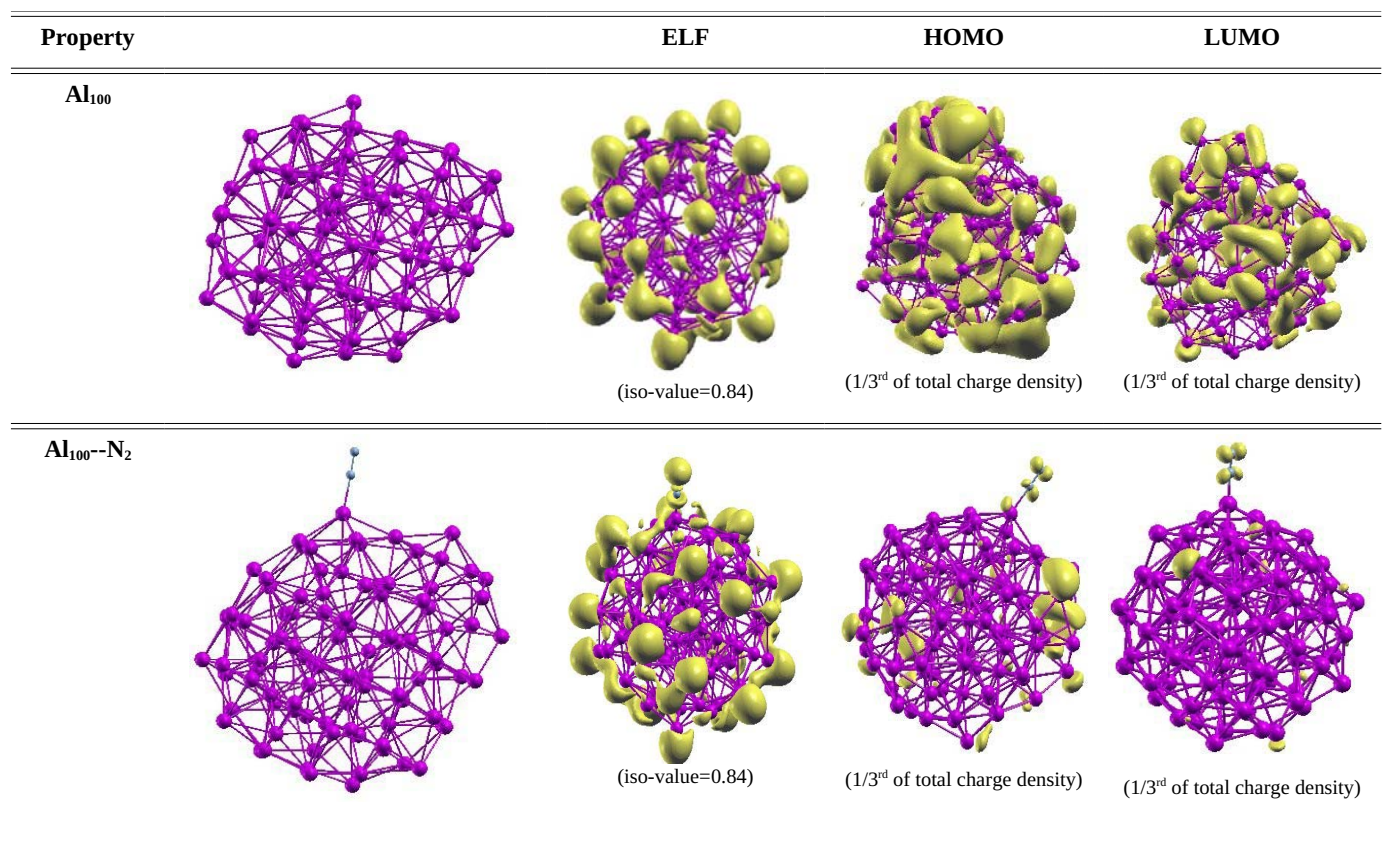

| Property | ELF | HOMO | LUMO |
|---|---|---|---|
| Al$_{100}$ | (iso-value=0.84) | (1/3$^{rd}$ of total charge density) | (1/3$^{rd}$ of total charge density) |
| Al$_{100}$--N$_2$ | (iso-value=0.84) | (1/3$^{rd}$ of total charge density) | (1/3$^{rd}$ of total charge density) |

**Figure 6: Al$_{13}$ (D$_h$) and its N$_2$ interaction from ELF and FMO**

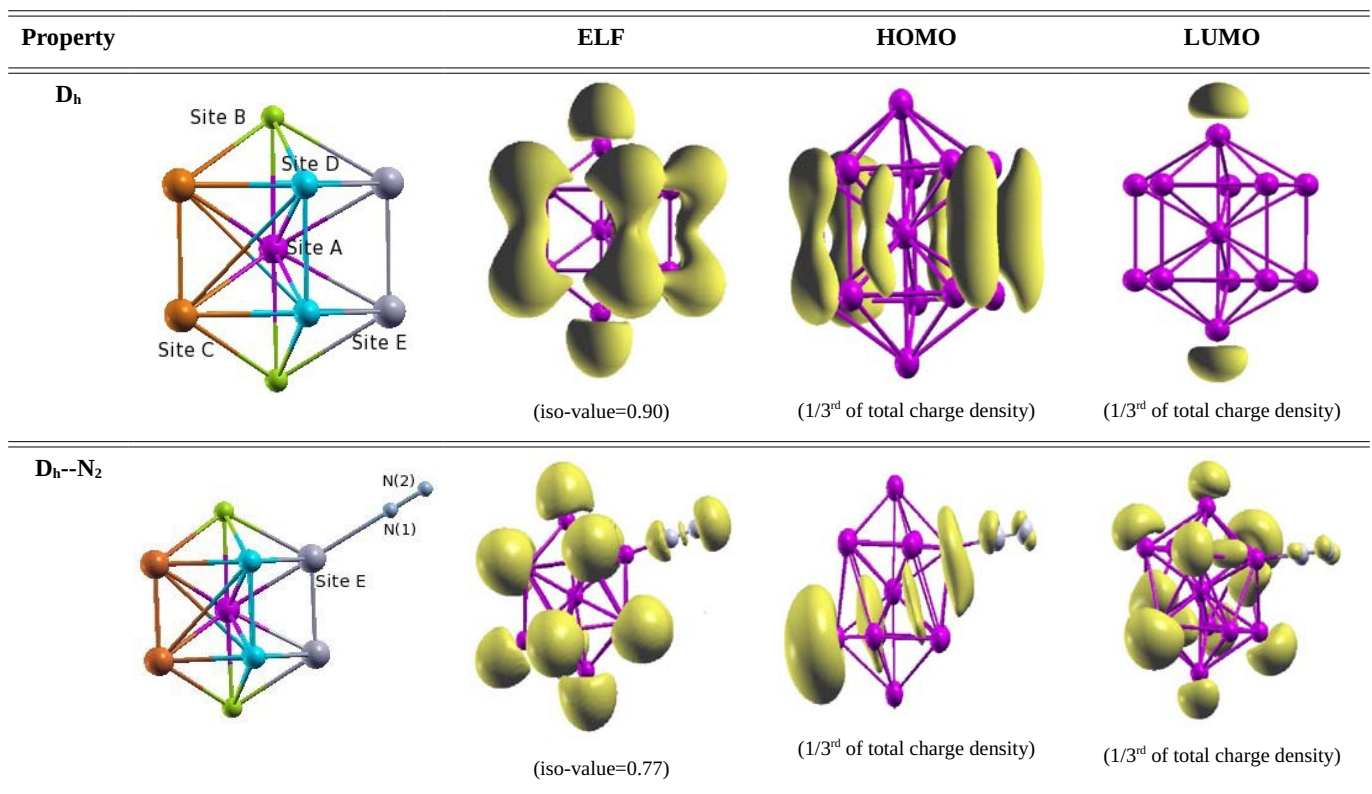

**Figure 7: Al$_{13}$ high energy configurations and its N$_2$ interaction from ELF and FMO**

| Property | | ELF | HOMO | LUMO |
|---|---|---|---|---|
| **H1** | | (iso-value=0.90) | (1/3$^{rd}$ of total charge density) | (1/3$^{rd}$ of total charge density) |
| **H1-N$_2$** | | (iso-value=0.90) | (1/3$^{rd}$ of total charge density) | (1/3$^{rd}$ of total charge density) |

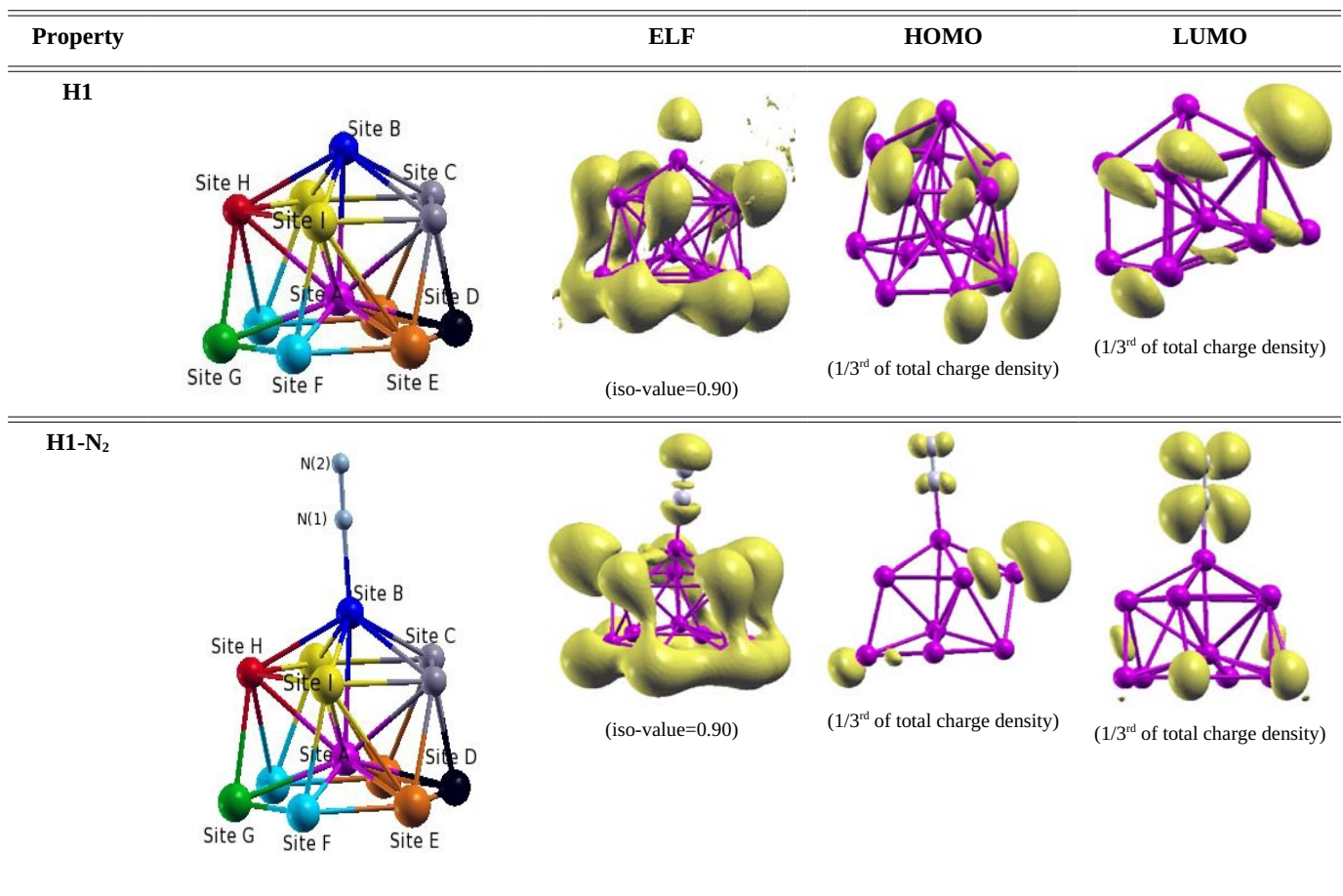

**H2**

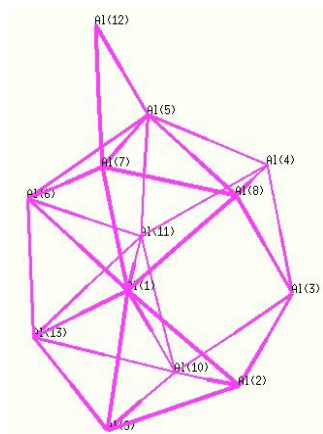
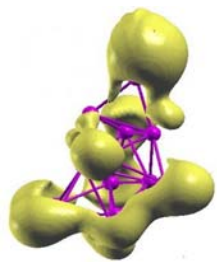

(iso-value=0.90)

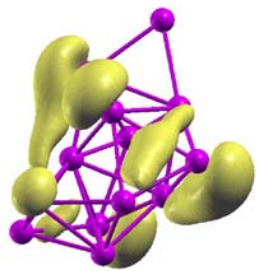

(1/3$^{rd}$ of total charge density)

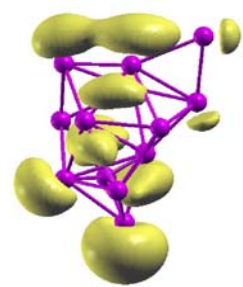

(1/3$^{rd}$ of total charge density)

**H2--N$_2$**

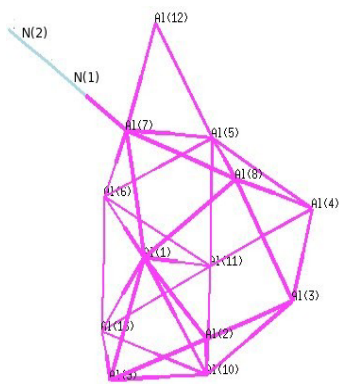
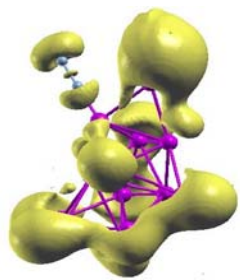

(iso-value=0.90)

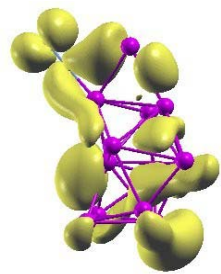

(1/3$^{rd}$ of total charge density)

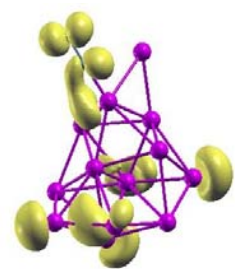

(1/3$^{rd}$ of total charge density)

**Figure 8: Interaction energy as a function of cluster size**

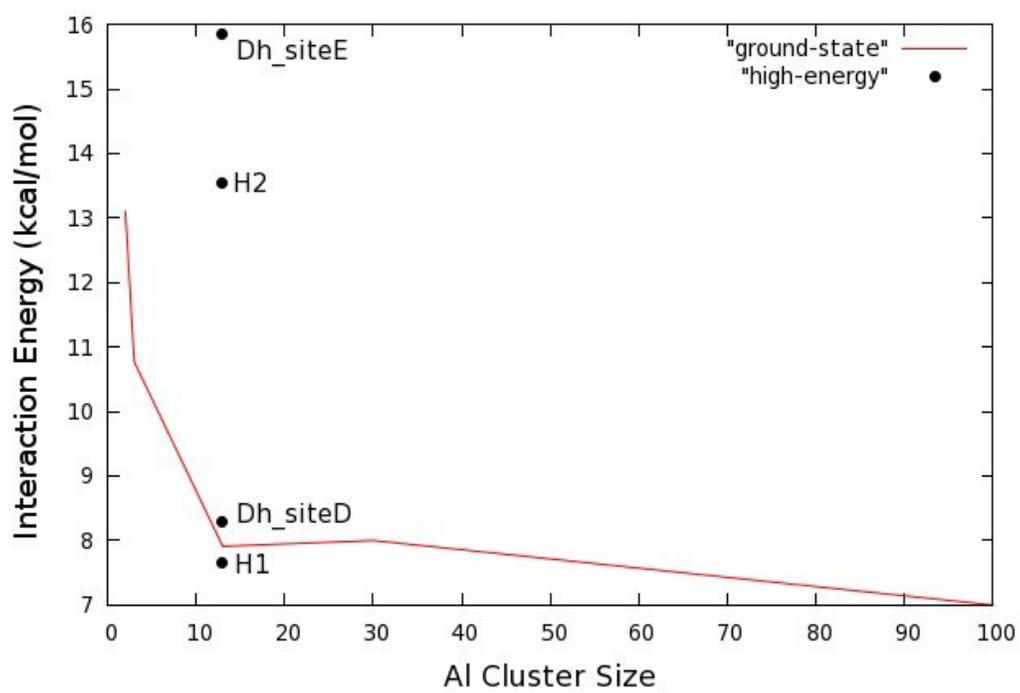